\documentclass{ifacconf}

\usepackage{graphicx}      
\usepackage{natbib}        
\usepackage{amsmath, bm,amsfonts,dsfont} 
\usepackage{verbatim} 
\usepackage{enumerate}
\usepackage{bbold}
\usepackage[table,xcdraw]{xcolor}
\definecolor{mooiblauw}{rgb}{0.25, 0.0, 1.0}

  {
      \theoremstyle{plain}
      \newtheorem{assumption}{Assumption}
  }
\providecommand{\openbox}{\leavevmode
  \hbox to.77778em{%
  \hfil\vrule
  \vbox to.675em{\hrule width.6em\vfil\hrule}%
  \vrule\hfil}}
\makeatletter
\DeclareRobustCommand{\qed}{%
  \ifmmode
    \eqno \def\@badmath{$$}
    \let\eqno\relax \let\leqno\relax \let\veqno\relax
    \hbox{\openbox}%
  \else
    \leavevmode\unskip\penalty9999 \hbox{}\nobreak\hfill
    \quad\hbox{\openbox}%
  \fi
}
\makeatother
\begin{document}
\begin{frontmatter}


\title{Exerting Control in Repeated Social Dilemmas with Thresholds\thanksref{footnoteinfo}}

\thanks[footnoteinfo]{The work was supported in part by the European Research Council (ERC-CoG-771687) and the Netherlands Organization for Scientific Research (NWO-vidi-14134).}

\author[First]{K. Frieswijk} 
\author[First]{A. Govaert} 
\author[First]{M. Cao}

\address[First]{ENTEG, Faculty of Science and Engineering, University of Groningen, The Netherlands, \\
(e-mail: \{k.frieswijk, a.govaert, m.cao\}@rug.nl)}

\begin{abstract}                
Situations in which immediate self-interest and long-term collective interest conflict often require some form of influence to prevent them from leading to undesirable or unsustainable outcomes. Next to sanctioning, social influence and social structure, it is possible that strategic solutions can exist for these social dilemmas. However, the existence of strategies that enable a player to exert control in the long-run outcomes can be difficult to show and different situations allow for different levels of strategic influence.
Here, we investigate the effect of threshold nonlinearities on the possibilities of exerting unilateral control in finitely repeated $n$-player public goods games and snowdrift games.
These models can describe situations in which a collective effort is necessary in order for a benefit to be created. We identify conditions in terms of a cooperator threshold for the existence of generous, extortionate and equalizing zero-determinant (ZD) strategies. 
Our results show that, for both games, the thresholds prevent equalizing ZD strategies from existing. 
In the snowdrift game, introducing a cooperator threshold has no effect on the region of feasible extortionate ZD strategies. 
For extortionate strategies in the public goods game, the threshold only restricts the region of enforceable strategies for small values of the public goods multiplier.
Generous ZD strategies exist for both games, but introducing a cooperator threshold forces the slope more towards the value of a fair strategy, where the player has approximately the same payoff as the average payoff of his opponents. 
\end{abstract}

\begin{keyword} game theory, repeated games, multiplayer games, ZD strategies.
\end{keyword}

\end{frontmatter}

\section{Introduction}

Social dilemmas arise when immediate self-interests conflict with long-term collective interests [\cite{van2013psychology}]. 
In these situations, selfish and myopic decisions can easily lead to undesirable collective outcomes; a situation that is most effectively described by the \textit{tragedy of the commons} [\cite{hardin1968tragedy}]. 
Social dilemmas exist in all sorts and sizes, ranging from global ecological concerns such as over-fishing and climate change to the social dilemma of autonomous vehicles [\cite{bonnefon2016social}].
Fortunately, not all social dilemmas collapse into unsustainable or undesirable outcomes. 
Many examples can be found in which a scenario like the tragedy of the commons is averted through complex mechanisms that affect economic, social and evolutionary decision-making processes. 
Research aimed at identifying these means of solving social dilemmas dates back decades [\cite{hardin1971collective,ostrom1990governing,hamilton1964genetical}], but remains relevant today [\cite{hauser2014cooperating,rand2013human,hilbe2018partners,bonnefon2016social}] and has identified a variety of solutions. A review discussing control of evolutionary games can be found in [\cite{Mingwildezeerin}]. 

However, many of these solutions are built upon restricting assumptions regarding the employed decision-making strategies.
Folk theorems, for instance, rely on rationality principles that can be violated when individuals make irrational decisions due to mistakes, fairness or spite.
Other solutions, like network reciprocity [\cite{nowak2006five}], through which cooperation can be sustained via a social network structure, do not immediately rely on the way decisions are made, but their effectiveness in solving social dilemmas can be affected significantly. 

In reality, one often does not know the precise decision-making trade-offs of individuals and what type of strategies they employ.  
This motivates the development of solutions to social dilemmas that are robust with respect to variation in behaviours. Recently, strategic solutions were identified in which a player, or a small group of players, can \textit{unilaterally} exert influence in the long-run outcome of social dilemmas [\cite{Press10409,Hilbe16425}]. 
Like the solution mechanisms that came before, these strategic solutions, known as \textit{zero-determinant (ZD) strategies}, also have their downsides. 
For one, they are defined over an \textit{infinite} horizon of repeated interactions, that has to be addressed using discounting methods [\cite{ichinose2018zero,hilbe2015partners,Govaert}]. Secondly, their existence can be challenging to show, and can be lost when there is no strict hierarchy in behaviour, like in Rock-Paper-Scissors games [\cite{stewart2016evolutionary}].
The \textit{identification} of classes of games that allow this form of strategic influence, just like the identification of potential games [\cite{monderer1996potential}], thus becomes an important problem. 
After existence has been shown, it is often not immediately clear how \textit{varied} of an influence can be exerted.
In general, this depends on the structure of the social dilemma, and thus different settings (e.g. public goods games, volunteers dilemma, snowdrift games etc.) give rise to different possibilities for exerting control. 
But even within a particular game, the level of control is affected by parameter values such as the benefit-to-cost ratio and group sizes. 

In this paper, we study $n$-player repeated social dilemmas with a \textit{finite} number of expected interactions. To provide a clearer specification of the types of social dilemmas that allow for the exertion of unilateral control, we introduce \textit{threshold nonlinearities}, and explore their influence on the existence of generous, extortionate and equalizer ZD strategies. Thresholds in the payoffs of social dilemmas are common in the literature [\cite{santos2011risk,pacheco2008evolutionary,souza2009evolution,liang2015analysis,hauser2014cooperating}] and are motivated by situations in which a collective effort is required to generate a benefit.
One can for instance think of obstacles that can only be removed by a mutual effort, or simply of an investment that requires a minimum of collective investment to become profitable.
The first scenario can be modelled by a \textit{threshold snowdrift game} [\cite{santos2012dynamics}], and the second by a \textit{threshold public goods game} [\cite{santos2011risk}]. 
Using the characterisation of enforceable payoff relations in [\cite{Govaert}], we investigate the often nonlinear relations between a variable threshold requirement, group-size and benefit-to-cost ratio on the level of strategic influence a player can exert.

This paper is organised as follows. In Section \ref{Preliminaries}, we introduce ZD strategies and present the assumptions made throughout the paper. In Section \ref{PGG}, we explore the existence of generous, extortionate and equalizing ZD strategies in the finitely repeated $n$-player linear public goods game. In Section \ref{SD}, we do the same, but then for the finitely repeated $n$-player snowdrift game. 

\section{Zero-determinant strategies and control in social dilemmas}\label{Preliminaries}

\noindent We consider symmetric social dilemmas in which players can cooperate (C) or defect (D).
The payoffs for the cooperators and defectors are given by $a_z$ and $b_z$, respectively, where $z$ denotes the number of cooperators among the co-players. For a social dilemma, we make the following natural assumptions~[\cite{Hilbe16425}]. 
\begin{assumption} (Social dilemma assumptions). The payoffs satisfy:
\begin{enumerate}[(i)]
    \item For all $0 \le z < n-1$, $a_{z+1} \ge a_z$ and $b_{z+1} \ge b_z$;
    \item For all $0 \le z < n-1$, $b_{z+1}>a_z$;
    \item $a_{n-1}> b_0$.
\end{enumerate}
\end{assumption}

Note that the above assumption implies that (i) each player prefers the other players to cooperate, irrespective of the player's own strategy; (ii) in any mixed group, defectors receive a strictly higher payoff; (iii) collective cooperation is favoured over collective defection.\\
In a repeated game, these single-round payoffs are averaged and discounted over the course of play. To denote the average discounted payoff compactly we introduce some notation. Define $$
  g^i:=(a_{n-1},\dots,a_0,b_{n-1},\dots,b_0 ) \in \mathbb{R}^{2n}  
$$
as the vector containing all possible payoffs of player $i$ in a given round of play.
Similarly, let
$g^{-i}_{C,z}:=\frac{za_{z}+(n-z-1)b_{z+1}}{n-1}$ and $g^{-i}_{D,z}:=\frac{za_{z-1}+(n-z-1)b_{z}}{n-1}$ denote the average payoff of $i$'s co-players given player $i$'s outcome $\{C,z\}$ and $\{D,z\}$, respectively.
Now define $$g^{-i}:=\left(g^{-i}_{C,n-1}, \hdots,g^{-i}_{C,0}, g^{-i}_{D,n-1} , \hdots, g^{-i}_{D,0}\right) \in \mathbb{R}^{2n}.$$



When future payoffs are discounted using an exponential discrete-time discounting function with a common and fixed discount factor $0<\delta<1$, the long-run expected payoff of player $i$ reads as [\cite{fudenberg1991tirole}]
\begin{equation}
   \pi^i=(1-\delta)\sum_{t=0}^\infty \delta^tg^i\cdot v(t),
\end{equation}
where $v(t)\in[0,1]^{2n}$ is the vector of outcome probabilities at time $t$. ZD strategies are memory-one strategies, which implies that they only take into account the outcome of the previous round [\cite{Press10409}]. 
Let $p_{x,{z}}$ denote player $i$'s conditional probability to cooperate, given that in the previous round, $i$ played $x\in\{C,D\}$ and $z$ co-players cooperated. Let us define
$$\mathbf{p}:=(p_{C,{n-1}},\dots,p_{C,0},p_{D,n-1},\dots,p_{D,0}).$$

Let $\mathbf{p}^{\textrm{rep}}:=(\mathds{1}_n,\mathbb{0}_n)$, and let $p_0\in[0,1]$ be player $i$'s initial probability to cooperate. In [\cite{Govaert}] it was shown that a ZD strategy of the form 
\begin{equation}\label{ZDrewritten}
\begin{aligned}
\delta\mathbf{p}=\mathbf{p}^{\textrm{rep}}+\phi\left[sg^i -g^{-i}+(1-s)l\mathds{1}\right]-(1-\delta)p_0\mathds{1},
\end{aligned}
\end{equation}
under the conditions that $\phi>0$, can enforce a linear relation in the average discounted payoffs, i.e.,
\begin{equation}
    \pi^{-i} = s \pi^i +(1-s)l.
\end{equation}

\noindent Here, $\pi^{-i} = \frac{1}{n-1}\sum_{j \neq i}^n \pi^j$. Table \ref{The one and only} summarizes the most studied linear payoff relations, and their respective strategies.

\begin{table}[ht]
\begin{center}
\caption{The four ZD strategies and their enforced linear payoff relation.}\label{The one and only}
\begin{tabular}{ccc}
\rowcolor[HTML]{EFEFEF} 
ZD-Strategy  & Parameter values    & Enforced payoff relation \\ \hline
Fair         & $s=1$               & $\pi^{-i}=\pi^i$         \\
Generous     & $l=a_{n-1}$, $0<s<1$ & $\pi^{-i}\geq\pi^i$      \\
Extortionate & $l=b_0$, $0<s<1$    & $\pi^{-i}\leq\pi^i$      \\
Equalizer    & $s=0$               & $\pi^{-i}=l$             \\ \hline
\end{tabular}
\end{center}
\end{table}

\noindent The baseline payoff $l$ has to satisfy [\cite{Govaert}]
\begin{equation}\label{l_inequalities}
   \begin{aligned}
l &\ge \max_{0 \le z \le n-1} \left\{ b_z - \tfrac{z}{n-1}\tfrac{b_z-a_{z-1}}{1-s} \right\}, \\
l &\le \min_{0 \le z \le n-1} \left\{ a_z + \tfrac{n-z-1}{n-1}\tfrac{b_{z+1}-a_{z}}{1-s} \right\},
\end{aligned} 
\end{equation}
\noindent with at least one strict inequality in \eqref{l_inequalities}. Moreover, for a finitely repeated $n$-player game, it is required that the slope of the linear payoff relation $s$ satisfies
\begin{align*}
    -\tfrac{1}{n-1}<& s<1,
\end{align*}

\noindent implying that there do not exist fair ZD strategies, for which $s=1$, in repeated $n$-player social dilemmas with a finite number of expected rounds.\\

\begin{figure*}
\centering
\includegraphics[width=0.85\linewidth]{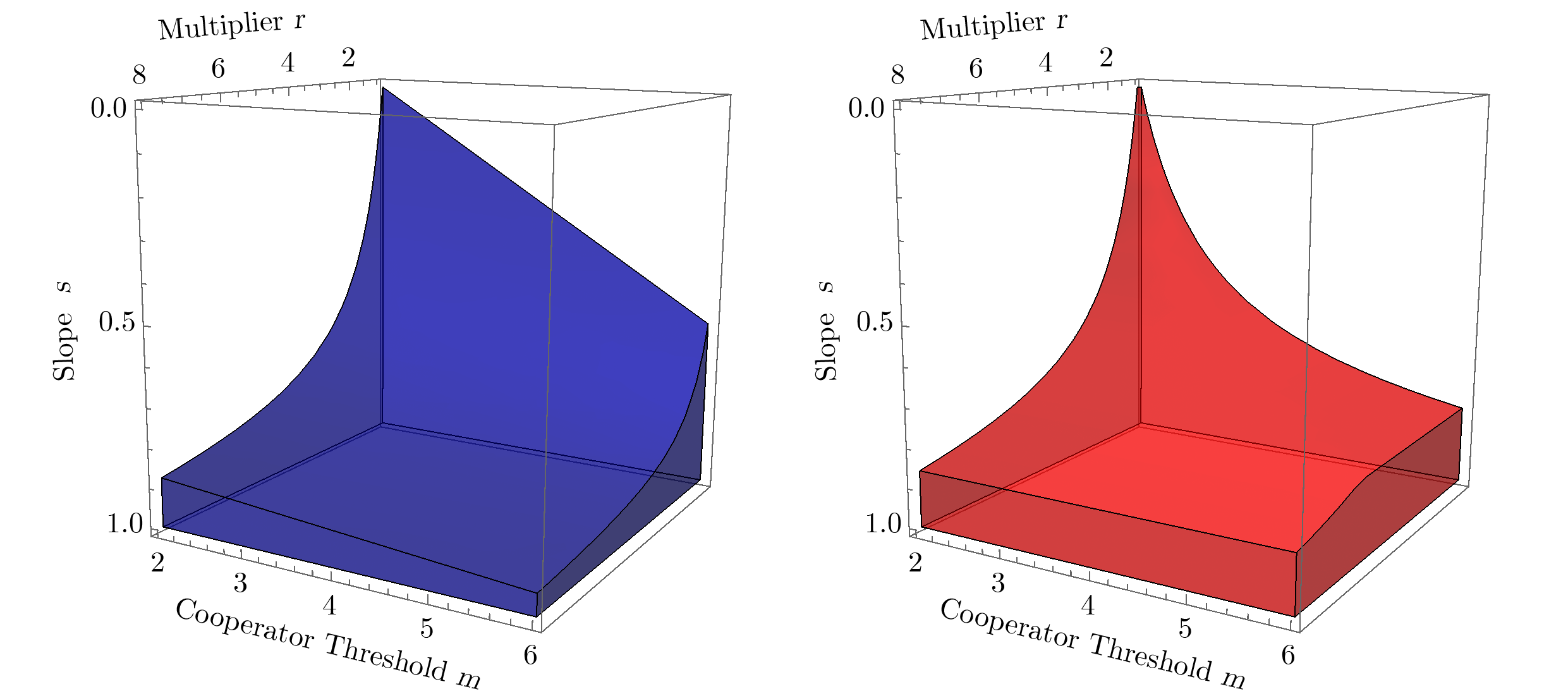}
\caption{Regions of strategy-existence in an $8$-player linear public goods game with a cooperator threshold $m$, for: generous strategies (left); extortionate strategies (right).} \label{fig:PGG}
\end{figure*}

\noindent Using the above, we are now ready to formulate the main goal of this paper, which is to identify conditions in terms of the cooperator threshold $m$ for the existence of generous, extortionate and equalizing ZD strategies.

\section{\lowercase{$n$}-player threshold public goods games} \label{PGG}

\noindent In this section, we explore the existence of generous, extortionate and equalizing strategies in the finitely repeated $n$-player threshold public goods game (PGG). The $n$ players either cooperate and invest $c>0$ into a public pot, or defect and invest nothing. If the total number of cooperators is greater than or equal to the threshold $m$, where $1<m < n$, the total sum of investments is multiplied by public goods multiplier $1<r<n$ and evenly divided among all players. If the total number of cooperators is lower than $m$, nobody receives any benefit. The payoffs of cooperators and defectors are thus given by~
\begin{alignat}{2} \label{PGGpayoffs}
\begin{split}
&a_z = \begin{cases}
         \ \tfrac{rc(z+1)}{n} -c & \text{if}\ z \ge m-1,\\
    \  -c & \text{if}\ z<m-1,     \end{cases}\\
& b_z = \begin{cases}
        \ \tfrac{rcz}{n}  \hspace{1.2cm} & \text{if}\ z \ge m,\\
     \ 0 \qquad& \text{if}\ z<m,
    \end{cases}
    \end{split}
\end{alignat}

\noindent respectively. By plugging the payoff functions \eqref{PGGpayoffs} into \eqref{l_inequalities}, we are able to derive the conditions under which generous, extortionate and equalizing ZD strategies exist. The results are presented in Theorem \ref{PGGtheorem}.

\begin{thm}\label{PGGtheorem}
Consider a public goods game with $1<r<n$, and payoffs \eqref{PGGpayoffs}. Then the following hold:
\begin{enumerate}[(i)]
    \item If a ZD strategy is \textit{generous}, i.e.\ if $l=a_{n-1}=rc-c$, and $0<s<1$, then every slope
    \begin{equation*}
s \ge 1- \dfrac{n-m+1}{r(n-1)}
\end{equation*}
\noindent can be enforced;
    \item If a ZD strategy is \textit{extortionate}, i.e.\ if $l=b_{0} =0$, and $0<s<1$, then every slope
\begin{equation*}
s \ge \max\left\{  \dfrac{m-2}{n-1}+ \epsilon, \ 1-\dfrac{n}{r(n-1)}\right\}
\end{equation*}
\noindent can be enforced,  where $\epsilon>0$ is an infinitesimally small number; 
    \item There do not exist \textit{equalizing} ZD strategies, i.e.\ there do not exist ZD strategies with $s=0$.
\end{enumerate}
\end{thm}

\noindent The proof of Theorem \ref{PGGtheorem} can be found in Appendix \ref{appPGG}. Theorem \ref{PGGtheorem} shows the influence of the cooperator threshold $m$ on the existence of ZD strategies. Observe that the range of enforceable slopes decreases for generous strategies as $m$ increases. For extortionate strategies on the other hand, the region of enforceable slopes only depends on $m$ if $r<\tfrac{n}{n-m+1}$. For $r \ge \tfrac{n}{n-m+1}$, the region of enforceable slopes is independent of $m$. Figure \ref{fig:PGG} depicts the regions of strategy-existence for a $n$-player public goods game, with $n=8$, for generous and extortionate strategies.

\section{\lowercase{$n$}-player snowdrift games}\label{SD}

\begin{figure*}
\centering
\includegraphics[width=0.85\linewidth]{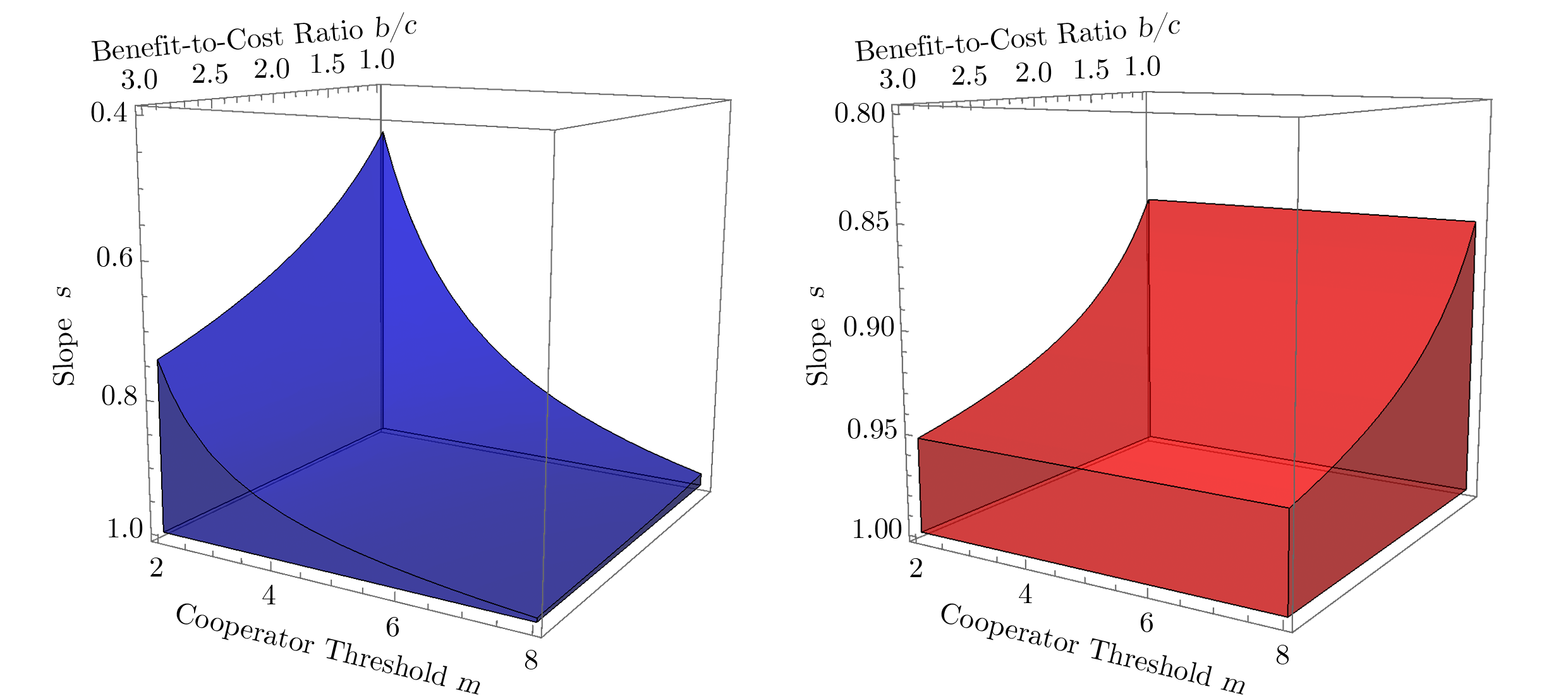}
\caption{Regions of strategy-existence in an $8$-player snowdrift game with a cooperator threshold $m$, for: generous strategies (left); extortionate strategies (right).}
\label{fig:SDG}
\end{figure*}

\noindent In the $n$-player snowdrift game (SDG), $n$ players get stuck in a snowdrift. In order to carry on, the snowdrift needs to be cleared. Each player has the choice to cooperate, and start shovelling, or defect, and do nothing. The cost of clearing the snowdrift, $c$, is shared by all cooperators. If the snowdrift is cleared, everyone obtains a benefit $b$, where $b>c>0$.\\
\noindent In contrast with the traditional snowdrift game, the amount of snow is so vast that there need to be at least $m$ cooperators in order to clear it, where $1<m<n$. The payoffs of cooperators and defectors are thus given by
\begin{alignat}{2} \label{SDpayoffs}
\begin{split}
&a_z = \begin{cases}
         \ b- \tfrac{c}{z+1}& \text{if}\ z \ge m-1,\\ 
    \  - \tfrac{c}{z+1} & \text{if}\ z<m-1, \end{cases}\\
& b_z = \begin{cases}
        \ b   \hspace{1.06cm} & \text{if}\ z \ge m,\\
     \ 0 & \text{if}\ z<m, 
    \end{cases}
    \end{split}
\end{alignat}
\noindent respectively. 
By plugging the payoff functions \eqref{SDpayoffs} into \eqref{l_inequalities}, we are able to derive the regions of existence for generous, extortionate and equalizing strategies. The results are presented in Theorem \ref{SDtheorem}. 

\begin{thm}\label{SDtheorem}
Consider a snowdrift game with $b>c>0$, and payoffs \eqref{SDpayoffs}. Then the following hold:
\begin{enumerate}[(i)]
    \item If a ZD strategy is \textit{generous}, i.e.\ if $l=a_{n-1}=b-\tfrac{c}{n}$, and $0<s<1$, then every slope
   \begin{equation*}
s \ge 1 - \dfrac{cn(n-m+1)}{(n-1)\big((bn-c)(m-1)+cn\big)}
\end{equation*}
\noindent can be enforced;
    \item If a ZD strategy is \textit{extortionate}, i.e.\ if $l=b_{0} =0$, and $0<s<1$, then every slope
\begin{equation*}
s \ge  1-\dfrac{c}{b(n-1)}
\end{equation*}
\noindent can be enforced; 
    \item There do not exist \textit{equalizing} ZD strategies, i.e.\ there do not exist ZD strategies with $s=0$.
\end{enumerate}
\end{thm}

The proof of Theorem \ref{SDtheorem} can be found in Appendix \ref{appSD}. Theorem \ref{SDtheorem} shows the influence of the threshold $m$ on the existence of ZD strategies. Figure \ref{fig:SDG} depicts the regions of strategy-existence for a $n$-player snowdrift game, with $n=8$, for generous and extortionate strategies.\\
Remarkably, for extortionate strategies, the region of enforceable slopes does not depend on the cooperator threshold $m$. For generous strategies, however, the value of $m$ matters a lot.  It can be observed from Figure \ref{fig:SDG} (left) that the higher the value of $m$ is, the higher the value of the slope $s$ has to be, which approaches 1 for higher values of $m$.

\section{Concluding Remarks}
\noindent We have studied repeated $n$-player public goods games and snowdrift games with a finite number of expected rounds, where we introduced a cooperator threshold $m$. We discovered that equalizing ZD strategies do not exist for these games when a cooperator threshold is imposed. We showed that in the snowdrift game, introducing a cooperator threshold has no effect on the region of feasible extortionate ZD strategies. For extortionate strategies in the public goods game, the threshold restricts the region of enforceable strategies only for small values of  public goods multiplier $r$. We observed that the threshold has a significant impact on the existence of generous ZD strategies. In particular, generous ZD strategies exist for both games, but a higher cooperator threshold forces the slope more towards the value of an approximately fair strategy, where the player's payoff is approximately equal to the average payoff of its opponents. \\
For future research, it would be interesting to explore the effect of the cooperator threshold on the minimum discount factor ($\delta$) that enables enforcing extortionate and generous payoff relations, shown in Table \ref{The one and only}.

\bibliography{bibliography}  

\appendix


\section{Analysis of \lowercase{$n$}-player linear public goods games} \label{appPGG} 

\noindent Consider a $n$-player public goods game with $1<r<n$, and $1<m<n$. Plugging the payoffs \eqref{PGGpayoffs} into \eqref{l_inequalities}, easily gives us the following results. For $z<m-1$, we obtain 
\begin{equation}\label{PGGz<m-1}
0 \le l \le \min \left\{ -c + \tfrac{(n-m+1)c}{(n-1)(1-s)}, rc-c \right\}.
\end{equation}

\noindent For $z=m-1$, we find
\begin{equation}\label{PGGz=m-1}
0 \le l \le \min \left\{ \tfrac{rcm}{n}-c + \tfrac{(n-m)c}{(n-1)(1-s)}, rc-c \right\}.
\end{equation}

\noindent For $z\ge m$, we have
\begin{equation}\label{PGGz>m}
\begin{aligned}
 l & \ge \max \left\{ \tfrac{rc(n-1)}{n}- \tfrac{c}{1-s}, 0 \right\},\\
 l & \le \min \left\{ \tfrac{rc(m+1)}{n}-c + \tfrac{(n-m-1)c}{(n-1)(1-s)}, rc-c \right\}.
\end{aligned}
\end{equation}

\noindent By using \eqref{PGGz<m-1}, \eqref{PGGz=m-1}, and \eqref{PGGz>m}, we can prove Theorem \ref{PGGtheorem}.
\subsection{Proof of Theorem \ref{PGGtheorem} (i)}\label{appPGGgen}
\noindent For a generous ZD strategy, we have parameter values  $l=a_{n-1}=rc-c$, and $0<s<1$. \\
\noindent For $z<m-1$, \eqref{PGGz<m-1} gives
\noindent 
\begin{equation*}
0 \le rc-c \le \min \left\{ -c + \tfrac{(n-m+1)c}{(n-1)(1-s)}, rc-c \right\}.
\end{equation*}
\noindent In order for generous strategies to exist, we must have
\begin{equation*}
-c + \tfrac{(n-m+1)c}{(n-1)(1-s)} \ge rc-c,
\end{equation*}
\noindent or equivalently $s \ge 1 - \tfrac{n-m+1}{r(n-1)}$. 
\noindent For $z=m-1$, \eqref{PGGz=m-1} yields
\begin{equation*}
0 \le rc-c \le \min \left\{ \tfrac{rcm}{n}-c + \tfrac{(n-m)c}{(n-1)(1-s)}, rc-c \right\}.
\end{equation*}
\noindent In order for generous strategies to exist, we must have
\begin{equation*}
\tfrac{rcm}{n}-c + \tfrac{(n-m)c}{(n-1)(1-s)} \ge rc-c,
\end{equation*}
\noindent or equivalently
\begin{equation*}
s \ge 1 - \tfrac{n}{r(n-1)}.
\end{equation*}

\noindent For $z \ge m$, \eqref{PGGz>m} gives
\begin{align}
 rc-c & \ge \max \left\{ \tfrac{rc(n-1)}{n}- \tfrac{c}{1-s}, 0 \right\} \label{PGG_RHSzgem},\\
 rc-c & \le \min \left\{ \tfrac{rc(m+1)}{n}-c + \tfrac{(n-m-1)c}{(n-1)(1-s)}, rc-c \right\}. \notag
\end{align}
\noindent In order for generous strategies to exist, we must have
\begin{equation*}
\tfrac{rc(m+1)}{n}-c + \tfrac{(n-m-1)c}{(n-1)(1-s)} \ge rc-c,
\end{equation*}
\noindent or equivalently $s \ge 1 - \tfrac{n}{r(n-1)}$. Note that if $s \ge 1 - \tfrac{n}{r(n-1)}$, then $\tfrac{rc(n-1)}{n}- \tfrac{c}{1-s} \le 0$, and \eqref{PGG_RHSzgem} becomes $rc-c \ge 0$.\\
Since $ 1 - \tfrac{n-m+1}{r(n-1)} > 1 - \tfrac{n}{r(n-1)}$, it follows that generous strategies exist for
\begin{equation*}
s \ge  1 - \tfrac{n-m+1}{r(n-1)}. \qed
\end{equation*}

\subsection{Proof of Theorem \ref{PGGtheorem} (ii)}\label{appPGGext}

\noindent For an extortionate ZD strategy, we have parameter values  $l=b_{0}=0$, and $0<s<1$. For $z<m-1$, \eqref{PGGz<m-1} gives
\begin{equation*}
0 \le 0 \le \min \left\{ -c + \tfrac{(n-m+1)c}{(n-1)(1-s)}, rc-c \right\}.
\end{equation*}
\noindent In order for extortionate strategies to exist, we must have
\begin{equation*}
-c + \tfrac{(n-m+1)c}{(n-1)(1-s)} \ge 0,
\end{equation*}
\noindent or equivalently $s \ge \tfrac{m-2}{n-1}$. Note that the fact that at least one of the $l$-inequalities in \eqref{l_inequalities} needs to be strict implies that $s > \tfrac{m-2}{n-1}$.\\
For $z=m-1$, \eqref{PGGz=m-1} gives
\begin{equation*}
0 \le 0 \le \min \left\{ \tfrac{rcm}{n}-c + \tfrac{(n-m)c}{(n-1)(1-s)}, rc-c \right\}.
\end{equation*}
\noindent Note that $s> -\tfrac{1}{n-1}$ implies that 
\begin{equation*}
\tfrac{rcm}{n}-c + \tfrac{(n-m)c}{(n-1)(1-s)} > 0,
\end{equation*}
\noindent so the lower bound does not exceed the upper bound. 

\noindent For $z \ge m$, \eqref{PGGz>m} gives
\begin{align*}
0 & \ge \max \left\{ \tfrac{rc(n-1)}{n}- \tfrac{c}{1-s}, 0 \right\},\\
0 & \le \min \left\{ \tfrac{rc(m+1)}{n}-c + \tfrac{(n-m-1)c}{(n-1)(1-s)}, rc-c \right\}.
\end{align*}

\noindent In order for extortionate strategies to exist, we must have
\begin{equation*}
 \tfrac{rc(n-1)}{n}- \tfrac{c}{1-s} \le 0,
\end{equation*}
\noindent or equivalently $s \ge 1 - \tfrac{n}{r(n-1)}$. \\
\noindent For $s \ge 1 - \tfrac{n}{r(n-1)}$, we have
\begin{equation*}
\tfrac{rc(m+1)}{n}-c + \tfrac{(n-m-1)c}{(n-1)(1-s)} \ge rc-c > 0,
\end{equation*}

\noindent which shows that the lower-bound does not exceed the upper-bound. Thus, extortionate strategies exist for
\begin{equation*}
s \ge \max\left\{  \tfrac{m-2}{n-1}+ \epsilon, \ 1-\tfrac{n}{r(n-1)}\right\}, 
\end{equation*}

\noindent with $\epsilon>0$ an infinitesimally small number. \qed

\subsection{Proof of Theorem \ref{PGGtheorem} (iii)}\label{appPGGeq}

\noindent For an equalizing ZD strategy, we have parameter value $s=0$. For $z<m-1$, \eqref{PGGz<m-1} gives
\begin{equation*}
0 \le l \le \min \left\{ -c + \tfrac{(n-m+1)c}{n-1}, rc-c \right\}.
\end{equation*}
\noindent In order for the lower bound to not exceed the upper bound, we must have 
\begin{equation*}
-c + \tfrac{(n-m+1)c}{n-1} \ge 0,
\end{equation*}
\noindent which implies $m\le 2$. Since $m>1$, it follows that $m=2$, for which the upper bound is equal to 0. However, since at least one of the $l$-inequalities in $0 \le l \le 0$ needs to be strict, there do not exist equalizing strategies. \qed

\section{Analysis of \lowercase{$n$}-player snowdrift games}  \label{appSD}

\noindent Consider a $n$-player snowdrift game with $b>c>0$, and $1<m<n$. Plugging the payoffs \eqref{SDpayoffs} into \eqref{l_inequalities} yields the following results. For $z<m-1$, we find
\begin{equation}\label{SDz<m-1}
0 \le l \le \min \left\{ \tfrac{c}{m-1}\left( \tfrac{n-m+1}{(n-1)(1-s)} -1\right), b-\tfrac{c}{n} \right\}.
\end{equation}

\noindent For $z=m-1$, we obtain
\begin{align*}
0 \le l & \le  \min \left\{ b- \tfrac{c}{m}+ \tfrac{(n-m)c}{m(n-1)(1-s)}, b-\tfrac{c}{n}  \right\} = b-\tfrac{c}{n}, 
\end{align*}
\noindent where we used $s> - \tfrac{1}{n-1}$. So
\begin{equation}\label{SDz=m-1}
0 \le l  \le  b-\tfrac{c}{n}.
\end{equation}

\noindent For $z\ge m$, we have
\begin{equation}\label{SDz>m}
\max \left\{ b - \tfrac{c}{(n-1)(1-s)}, 0 \right\} \le l \le b-\tfrac{c}{n}.
\end{equation}
\noindent By using \eqref{SDz<m-1}, \eqref{SDz=m-1}, and \eqref{SDz>m}, we can prove Theorem \ref{SDtheorem}.

\subsection{Proof of Theorem \ref{SDtheorem} (i)}\label{appSDgen}
For a generous ZD strategy, we have parameter values  $l=a_{n-1}=b-\tfrac{c}{n}$, and $0<s<1$. For $z<m-1$, \eqref{SDz<m-1} gives
\begin{equation*}
0 \le b-\tfrac{c}{n} \le \min \left\{ \tfrac{c}{m-1}\left( \tfrac{n-m+1}{(n-1)(1-s)} -1\right), b-\tfrac{c}{n} \right\}.
\end{equation*}
\noindent In order for generous strategies to exist, we must have
\begin{equation*}
\tfrac{c}{m-1}\left( \tfrac{n-m+1}{(n-1)(1-s)} -1\right) \ge b-\tfrac{c}{n},
\end{equation*}
\noindent or equivalently $s \ge 1 - \tfrac{cn(n-m+1)}{(n-1)\big((bn-c)(m-1)+cn\big)}$.\\
\noindent For $z=m-1$,  \eqref{SDz=m-1} gives
\begin{equation*}
0 \le b-\tfrac{c}{n} \le b-\tfrac{c}{n},
\end{equation*}
\noindent which is always satisfied.\\
\noindent For $z \ge m$, \eqref{SDz>m} gives
\begin{equation*}
\max \left\{ b - \tfrac{c}{(n-1)(1-s)}, 0 \right\} \le b-\tfrac{c}{n}  \le b-\tfrac{c}{n}.
\end{equation*}

\noindent If $s> -\tfrac{1}{n-1}$, then $ b - \tfrac{c}{(n-1)(1-s)} < b-\tfrac{c}{n} $, so the lower bound does not exceed the upper bound. Thus, generous strategies exist for
\begin{equation*}
s \ge 1 - \tfrac{cn(n-m+1)}{(n-1)\big((bn-c)(m-1)+cn\big)}. \qed
\end{equation*}

\subsection{Proof of Theorem \ref{SDtheorem} (ii)}\label{appSDext}

\noindent For an extortionate ZD strategy, we have parameter values  $l=b_{0}=0$, and $0<s<1$. For $z<m-1$,  \eqref{SDz<m-1} gives
\begin{equation*}
0 \le 0 \le \min \left\{ \tfrac{c}{m-1}\left( \tfrac{n-m+1}{(n-1)(1-s)} -1\right), b-\tfrac{c}{n} \right\}.
\end{equation*}

\noindent In order for extortionate strategies to exist, we must have
\begin{equation*}
\tfrac{c}{m-1}\left( \tfrac{n-m+1}{(n-1)(1-s)} -1\right) \ge 0,
\end{equation*}
\noindent or equivalently $s \ge \tfrac{m-2}{n-1}$. Note that the fact that at least one of the $l$-inequalities in \eqref{l_inequalities} needs to be strict implies that $s > \tfrac{m-2}{n-1}$.

\noindent For $z=m-1$,  \eqref{SDz=m-1} yields
\begin{equation*}
0 \le 0 \le  b-\tfrac{c}{n}.
\end{equation*}
\noindent Since $b>c>0$, we have $b-\tfrac{c}{n}>0$, so the lower bound does not exceed the upper bound.

\noindent For $z\ge m$,  \eqref{SDz>m} gives
\begin{equation*}
\max \left\{ b - \tfrac{c}{(n-1)(1-s)}, 0 \right\} \le 0 \le b-\tfrac{c}{n}.
\end{equation*}

\noindent In order for extortionate strategies to exist, we must have
\begin{equation*}
b - \tfrac{c}{(n-1)(1-s)} \le 0,
\end{equation*}
\noindent or equivalently $s \ge 1 - \tfrac{c}{b(n-1)}$. Since $b-\tfrac{c}{n}>0$, the lower bound does not exceed the upper bound for $s \ge 1 - \tfrac{c}{b(n-1)}$. Note that $b>c>0$ implies that $-\tfrac{c}{b} > -1$. Hence, $1 - \tfrac{c}{b(n-1)} > 1 - \tfrac{1}{n-1} = \tfrac{n-2}{n-1} > \tfrac{m-2}{n-1}$.

Thus, extortionate strategies exist if and only if
\begin{equation*}
s \ge     1-\tfrac{c}{b(n-1)}. \qed
\end{equation*}

\subsection{Proof of Theorem \ref{SDtheorem} (iii)}\label{appSDeq}

\noindent For an equalizing ZD strategy, we have parameter value $s=0$. For $z<m-1$, \eqref{SDz=m-1} gives
\begin{equation*}
0 \le l \le \min \left\{ \tfrac{c}{m-1}\left( \tfrac{n-m+1}{n-1} -1\right), b-\tfrac{c}{n} \right\}.
\end{equation*}

In order for the lower bound to not exceed the upper bound, we must have 
\begin{equation*}
\tfrac{c}{m-1}\left( \tfrac{n-m+1}{n-1} -1\right) \ge 0,
\end{equation*}
\noindent which implies $m\le 2$. Since $m>1$, it follows that $m=2$, for which the upper bound is equal to 0. However, since at least one of the $l$-inequalities in $0 \le l \le 0$ needs to be strict, there do not exist equalizing strategies. \qed

\end{document}